\documentstyle[preprint,aps,epsf]{revtex}                  
\global\let\epsfloaded=Y 
\begin{document}
\pagestyle{empty}                                      
\preprint{
\font\fortssbx=cmssbx10 scaled \magstep2
\hbox to \hsize{
\hbox{
}
\hfill $
\vtop{              
 \hbox{ }}$
}
}
\draft
\vfill
\title{
Direct $CP$ Violation in Angular Distribution of  \\
$B\to J/\psi K^{*}$ Decays
}

\vfill
\author{$^{1,2}$Xiao-Gang He and $^2$Wei-Shu Hou}
\address{
\rm $^1$ School of Physics, University of Melbourne, 
Parkville, Vic. 3052, Australia\\
\rm $^2$Department of Physics, National Taiwan University,
Taipei, Taiwan 10764, R.O.C.
}

%
%
\vfill
\maketitle
\begin{abstract}
We show that the study of 
certain observables in the 
angular distribution in $B\to J/\psi K^*$ provide clear test
for CP vioaltion beyond the Standard Model. 
These observables vanish in SM,
but in models beyond SM some of them can be 
large enough to be measured at B factories.
\end{abstract}

%
%
%
%
\pagestyle{plain}
The CLEO Collaboration has recently reported \cite{Pwave} 
the first full angular analysis of 
$B\to  J/\psi K^{*}$ decays.
Several CP conserving quantities have been studied.
They find that the $P$ wave component is small,
$\vert P\vert^2 = 0.16 \pm0.08\pm 0.04$, which is good news for measuring mixing induced 
CP violation via $B\to J/\psi K^{*0} \to J/\psi K_S \pi^0$ decay\cite{hh}. 
Relative to the longitudinal amplitude $A_0$, the phases
of the transverse and parallel amplitudes
were found to be $\phi(A_T)= -0.11\pm0.46\pm 0.03$ rad, and $\phi(A_{||})=3.00\pm 0.37\pm 
0.04$ rad, respectively. These phases are consistent with 0 or $\pi$, indicating 
the absence of final state interaction (FSI) phase shifts if CP is conserved.
It should be noted that the analysis in fact does not 
measure FSI phases alone, 
but a combination of FSI and CP violating weak phases. 
In order to have information about FSI and CP violating weak phases, 
the particle and anti-particle decays have to be separately measured.
With increased luminosities  in the near future at CLEO, the SLAC and KEK 
B-factories and other 
facilities,  together with good $K/\pi$ separation capabilties (particle identification), 
one can distinguish $K^*$ from $\bar K^*$, hence $B$ vs. $\bar B$. One can then  
study direct CP violation in these decays even if the partial rate asymmetry is
zero. In this paper we show that certain observables in the 
angular distribution in $B\to J/\psi K^*$ provide clear test
for CP violation beyond the Standard Model.

All necessary information for the present study is contained in the full angular 
distribution
for $B\to J/\psi K^*$ which is given by\cite{dunietz}
\begin{eqnarray}
&&{1\over \Gamma} {d^3\Gamma \over d\cos\theta_{\rm tr} d\cos \theta_{K^*}
d\phi_{\rm tr}}\nonumber\\
&&= {9\over 32\pi} \left\{2|A_0|^2 \cos^2\theta_{K^*}(1-\sin^2\theta_{\rm tr} \cos^2\phi_{\rm tr})
+ |A_{||}|^2 \sin^2\theta_{K^*} (1-\sin^2\theta_{\rm tr}\sin^2\phi_{\rm tr})
     \phantom{1\over \sqrt{2}}\right.\nonumber\\
&&+ |A_{T}|^2 \sin^2\theta_{K^*} \sin^2\theta_{\rm tr}
- {\rm Im}\, (A^*_{||} A_{T}) \, \sin^2\theta_{K^*} \sin2\theta_{\rm tr}\sin \phi_{\rm tr}\nonumber\\
&&  \left. + {1\over \sqrt{2}} {\rm Re}\, (A^*_0 A_{||})
           \, \sin2\theta_{K^*} \sin^2\theta_{\rm tr}\sin2\phi_{\rm tr}
       + {1\over \sqrt{2}} {\rm Im}\, (A^*_0A_{T})
           \, \sin2\theta_{K^*} \sin2\theta_{\rm tr}\cos\phi_{\rm tr}\right\},
\label{angular}
\end{eqnarray}
where $A_{T}=P$ is the $P$ wave decay amplitude, and $A_{0}$ and $A_{||}$ are 
two othorgonal combinations of  the $S$ and $D$ wave amplitudes with the
normalization $|A_{T}|^2 + |A_0|^2 + |A_{||}|^2 = 1$. 
The angles $\theta_{\rm tr}$ and
$\phi_{\rm tr}$ are defined as polar and azimuth angles of the charged lepton in the
$J/\psi$ rest frame, with $x$ axis along the direction of $K^*$, and $x$-$y$ plane 
parallel to $K\pi$ plane. The angle $\theta_{K^*}$ is defined as that of the 
$K$ in the rest frame of $K^*$ 
relative to the negative of the $J/\psi$ direction in that frame.
The angular distribution for $\bar B$ decay is similar, and we shall use $\bar A$ to indicate 
the corresponding amplitudes. 

In the CLEO analysis,
the phase for $A_0$ was taken to be zero. 
For convenience we will use the convention that 
each amplitude $A_i$ has both CP conserving FSI phase
$\phi_i$ and CP violating phase $\sigma_i$, 
i.e.  $A_{j} = |A_{j}|e^{i(\phi_{j} + \sigma_{j})}$ while
$\bar A_T = - |A_T|e^{i(\phi_T -\sigma_T)}$, 
$\bar A_{||} = |A_{||}|e^{i(\phi_{||} - \sigma_{||})}$, and
$\bar A_0 = |A_0|e^{i(\phi_0-\sigma_0)}$.
There are several ways for CP violation to manifest itself, 
the most familiar one is in partial rate asymmetries. 
Since there are three different decay amplitudes, 
partial rate asymmetries may show up in either of them\cite{gv}. 
These asymmetries can be studied by measuring the coefficients of 
the first three terms in Eq. (\ref{angular}) for $B$ and $\bar B$ decays 
and comparing them. 
However, in the case under consideration such differences are
very small and therefore difficult to measure.
It is therefore interesting to see if there are other observables in 
the angular distribution which provide useful information about
CP violation even if the partial rate asymmetries are zero.

It is clear that the coefficients 
$\alpha = -{\rm Im}\, (A_{||}^*A_T)$, $\beta = {\rm Re}\, (A_0^*A_{||})$, 
and $\gamma = {\rm Im}\, (A_0^*A_T)$
of the last three terms in the angular distribution,
and similarly $\bar \alpha$, $\bar \beta$ and $\bar \gamma$ for $\bar B$ decays,
contain information about CP violation.
Without separating $B$ and $\bar B$ decays,  however,
which was the case for the CLEO analysis mentioned earlier, 
information on CP violation cannot be extracted. 
One must obtain the angular distributions for 
$B\to J/\psi K^*$ and $\bar B\to J/\psi \bar K^*$ decays separately
and determine the coefficients for the interference terms in each case. 
The following three quantities then measure CP violation,
\begin{eqnarray}
 a_1= \alpha + \bar \alpha &=& \ \ 2|A_T||A_{||}|\cos(\phi_{||T})\sin(\sigma_{||T}),
\nonumber\\
a_2 = \beta - \bar \beta &=& -2|A_{||}||A_0|\sin(\phi_{||0})\sin(\sigma_{||0}),
\nonumber\\
a_3 = \gamma + \bar \gamma &=& -2|A_T||A_0|\cos(\phi_{T0})\sin(\sigma_{T0}).
\end{eqnarray}
where $\phi_{ij} = \phi_i-\phi_j$ and $\sigma_{ij} = \sigma_i -\sigma_j$.
Observables analogous to these, but integrated over various ranges of angles,
have already been considered in Ref.\cite{gv}.
We remark that similar analysis can be carried out for any two vector meson 
decay channels of $B$ meson, such as $B\to D^* \rho$ and $B\to \phi K^*$,
but it is probably more useful for tree dominated decays where the
rate asymmetry is small.

We note that the CP violating observables $a_{1,3}$ do not
require FSI phase differences and are especially sensitive to CP violating weak 
phases.
The present CLEO data on angular distributions
which provide information about FSI phases 
are proportional to the CP conserving quantities 
$\alpha - \bar \alpha \sim \sin(\phi_{||T})\cos(\sigma_{||T})$,
$\beta +\bar \beta \sim \cos(\phi_{||0})\cos(\sigma_{||0})$,
and $\gamma -\bar \gamma \sim \sin(\phi_{T0})\cos(\sigma_{T0})$.
At present these data do not exclude $a_{1}$ ($a_{3}$)  up to 0.50 (0.58), 
but the small FSI phase for $A_{||}$ measured by CLEO implies that $a_2$ is small.

In SM the decay amplitude for $B\to J/\psi K^*$ is due to 
the quark level effective Hamiltonian
\begin{eqnarray}
H_{\rm eff} = {G_F\over \sqrt{2}} V_{cb}V_{cs}^*
    \left\{C_1 \, \bar c \gamma_\mu (1-\gamma_5)c \, \bar s \gamma^\mu (1-\gamma_5) b
      +C_2 \, \bar s \gamma_\mu (1-\gamma_5) c \, \bar c \gamma^\mu (1-\gamma_5)b
    \right\},
\end{eqnarray}
where $C_1 = -0.313$ and $C_2=1.15$\cite{dh}, 
and we have neglected the negligibly small penguin contribution. 
This effective Hamiltonian generates a common weak phase for all the amplitudes through
the phase of $V_{cb}V_{cs}^*$, which is zero in the Wolfenstein 
parametrization.
The quantities $a_i$ therefore all vanish. However, in
extensions of SM, these phases need not be the same, 
and the values for $a_i$ may no longer vanish.  
Hence, the observables $a_i$ provide good tests for CP 
violation beyond SM.

There are many ways where new physics may change 
the phases $\sigma_i$. 
To lowest order they may arise from 
 dimension 6 four quark operators, or from the
dimension 5 color dipole operator $\bar s i\sigma_{\mu\nu}G^{\mu\nu} 
(1\pm \gamma_5) b$, where $G^{\mu\nu}$ is the gluon field strength.
New physics contributions 
from $C_{L} \, \bar c \gamma^\mu(1\pm \gamma_5)c \, \bar s \gamma_\mu (1-\gamma_5) b$
type of interactions are proportional to the SM contribution, which 
just generate a common weak phase for all the amplitudes and therefore the
$a_i$ are all zero.
The observables $a_i$ discussed here do not provide good tests
for new physics of this type.
Interactions of the form
$C_{R} \, \bar c\gamma^\mu (1\pm \gamma_5)c \, \bar s \gamma_\mu (1+\gamma_5) b$, 
however, will generate different phases for $A_T$ and $A_{||,0}$
because
the $\bar s \gamma^\mu(1+\gamma_5) b$ contribution to $A_T$ is proportional to
$C_R$, but for $A_{||,0}$ it is proportional to $-C_R$. To a good approximation,
one gets the weak phase $\sigma_T = - \sigma_{||,0}$, 
leading to non-zero values for $a_i$.

Let us consider, as an example, R-parity violating supersymmetric models. 
The exchange of charged sleptons or down type squarks can generate 
non-zero $C_R$ with an arbitrary phase $\sigma_R$. 
The allowed value for $C_R$ is constrained by 
experimental data on $b\to s\gamma$, but it still allows the $C_R$ 
contribution to the $B\to J/\psi K^*$ amplitude to be as large as
20\% of the SM contribution. 
Slightly stronger constraints can be obtained by
assuming that $b\to c \bar c  s$ is similar to $b\to c \bar u s$ from R-parity
violating interactions. 
The upper bound of the weak phases are approximately given by\cite{gw}
\begin{eqnarray}
\sigma_T=-\sigma_{||,0} \approx 0.1\sin \sigma_R.
\end{eqnarray}
Using the central values for the modulous of the amplitudes and assuming
that the FSI  phases are zero, we find 
\begin{eqnarray}
a_1 &\simeq& \ \ 0.10\sin\sigma_R\nonumber\\
a_3 &\simeq& -0.12\sin\sigma_R.
\end{eqnarray}

The contribution from the dimension five color dipole operator has been  
estimated by assuming that color octet operators 
contribute the amount as determined 
in generalized factorization approximation\cite{hh}. The magnitude of the color
dipole coefficient as large as 10 times that of the SM one is not ruled out. 
In fact, charm counting and semi-leptonic branching ratio problems 
in B decays\cite{ghk}, and perhaps the large $B\to \eta' X_s$, may 
favor such large value\cite{ht}.
The weak phases $\sigma_i$ can be as large as $0.1\sin\sigma_c$,
where $\sigma_c$ is the weak phase of the color dipole coupling.
However, if the new color dipole interaction has the same $\bar s\sigma_{\mu\nu}(1+\gamma_5)b$ 
chiral structure
as in SM, 
the phases are approximately equal and the values for $a_i$ would be 
very small. For a color dipole of 10 times the SM strength but with 
$\bar s\sigma_{\mu\nu}(1-\gamma_5)b$ chiral structure,
one obtains $\sigma_T \approx - \sigma_{||,0}$ leading to an upper bound
of $0.1\sin\sigma_c$. This would generate $a_1$ and $a_3$ 
as large as $0.1\sin\sigma_c$ and $-0.12\sin\sigma_c$ respectively,
quite similar to the R-parity violating case.

In all cases discussed above the weak phases for $\sigma_{||}$ and 
$\sigma_{0}$ are equal (or approximately equal).
The asymmetry $a_2$ is therefore approximately zero in all cases we have considered, and 
does not seem to be a good quantity to study for CP violation using this method.

The sensitivety for $a_{1,3}$ is similar to the sensitivity to the 
phase angles of the amplitudes.
It is interesting to note that the systematic error in the CLEO analysis  
is already as low as 0.03\cite{Pwave}.
With increased statistics, $a_{1,3}$ 
as large as 0.10 should be accessible 
at CLEO, at CDF, and at future B factories.
Since the errors are determined through a fit,
it is not clear how the statistical error scales with actual number of events.
The question can only be answered by actual studies,
but naive scaling implies that one would need $10^8$
events to be able to distinguish the deviations that we gave as illustration.
Nevertheless, the needed number of events may be less,
and measurement of the observables $a_i$ will provide us with 
useful information about CP violation.
We remark again that similar CP violating observables
can be constructed for any $B\to VV$ decays, such as
$B\to D^*\rho$, where preliminary evidence for FSI phases
has recently been reported by CLEO \cite{Palmer}.

In conclusion,
we have shown that 
the study of the observables $a_i$ in the angular distributions of 
$B\to J/\psi K^*$ and $\bar B\to J/\psi \bar K^*$
decays can provide good tests for CP violation beyond the Standard Model,
since they are  zero in  SM.
We have illustrated with supersymmetric models with R-parity violation, and models with 
large color dipole 
interactions, where these observables can be 
large enough to be measurable.

This work is supported in part by 
grant
NSC 87-2112-M-002-037 and NSC 87-2811-M-002-046 
of the Republic of China and by Australian Research Council.
WSH would like to thank the University fo Melbourne and
the Special Center for the Subatomic Structure of Matter at the
University of Adelaide for hospitality where part of this work was done.

\end{document}